\documentclass[%
 aip,
 amsmath,amssymb,
 reprint,%
]{revtex4-1}

\usepackage[english]{babel}
\usepackage{graphicx}
\usepackage{dcolumn}
\usepackage{bm}

\usepackage[utf8]{inputenc}
\usepackage[T1]{fontenc}
\usepackage{mathptmx}
\usepackage{etoolbox}
\usepackage{amsmath, amsfonts}
\usepackage{physics}
\usepackage{booktabs}
\usepackage[colorlinks=true,
            linkcolor=blue,
            urlcolor=blue,
            citecolor=blue]{hyperref}

\makeatletter
\def\@email#1#2{%
 \endgroup
 \patchcmd{\titleblock@produce}
  {\frontmatter@RRAPformat}
  {\frontmatter@RRAPformat{\produce@RRAP{*#1\href{mailto:#2}{#2}}}\frontmatter@RRAPformat}
  {}{}
}%
\makeatother
\begin{document}

\preprint{AIP/123-QED}

\title{Recurrent chaotic clustering and slow chaos in adaptive networks}
\author{Matheus Rolim Sales}
\email{rolim.sales@unesp.br}
\affiliation{Department of Physics, São Paulo State University, 13506-900, Rio Claro, SP, Brazil}
\affiliation{Graduate Program in Sciences, State University of Ponta Grossa, 84030-900, Ponta Grossa, PR, Brazil}
\affiliation{Potsdam Institute for Climate Impact Research, Member of the Leibniz Association, P.O. Box 6012 03, D-14412 Potsdam, Germany}
\author{Serhiy Yanchuk}
\affiliation{School of Mathematical Sciences, University College Cork, Western Road, Cork T12 XF62, Ireland}
\affiliation{Potsdam Institute for Climate Impact Research, Member of the Leibniz Association, P.O. Box 6012 03, D-14412 Potsdam, Germany}
\author{J\"urgen Kurths}
\affiliation{Potsdam Institute for Climate Impact Research, Member of the Leibniz Association, P.O. Box 6012 03, D-14412 Potsdam, Germany}
\affiliation{Institute of Physics, Humboldt University Berlin, 10099 Berlin, Germany}

\date{\today}

\begin{abstract}
Adaptive dynamical networks are network systems in which the structure co-evolves and interacts with the dynamical state of the nodes. We study an adaptive dynamical network in which the structure changes on a slower time scale relative to the fast dynamics of the nodes. We identify a phenomenon we refer to as recurrent adaptive chaotic clustering (RACC), in which chaos is observed on a slow time scale, while the fast time scale exhibits regular dynamics. Such slow chaos is further characterized by long (relative to the fast time scale) regimes of frequency clusters or frequency-synchronized dynamics, interrupted by fast jumps between these regimes. We also determine parameter values where the time intervals between jumps are chaotic and show that such a state is robust to changes in parameters and initial conditions.
\end{abstract}
\keywords{Adaptive networks, slow-fast dynamics, recurrent synchronization, slow chaos, cluster switching}
\maketitle

\begin{quotation}
Many complex dynamical systems have a network structure that evolve (adapt) over time. Such systems are typical in neural learning systems but also occur in a variety of other applications. A typical scenario is when the network structure changes much slower than the dynamics of each individual node. In such cases, many exciting multiscale phenomena occur that are impossible in systems with a static network structure. In this paper, we present the phenomenon of recurrent adaptive chaotic clustering (RACC) in which the network structure changes slowly and chaotically, while the node dynamics is fast and regular. In addition, the fast dynamics recurrently changes from complete frequency synchrony to frequency clusters of different types.
\end{quotation}

\section{Introduction}
\label{sec:intro}

Adaptive dynamical networks (ADN) appear in various applications ranging from neuroscience \cite{Abbott2000, Gerstner2014, Popovych2013} to social \cite{Gross2006,Gross2009,Horstmeyer2020,schweitzerSocialPercolationRevisited2021} or transportation \cite{Martens2019} networks. While dynamical networks with static connectivity can describe synchronization phenomena and pattern formation in networks with static structure \cite{Pikovsky2003a,reductionphase2}, ADNs allow a dynamic change of the network structure and an interaction between this structural dynamics and the dynamics of the network nodes \cite{Berner2023,gross2008adaptive}. ADNs are fundamental models for describing the learning of neuronal systems due to neuronal plasticity \cite{Caporale2008,Gerstner2014}. 

ADNs exhibit exciting new dynamical phenomena such as frequency clusters \cite{KAS17,Berner2019,Feketa2021,POP15,AOK15,Rohr2019}, recurrent synchronisation \cite{Thiele2023}, self-organised noise resistance \cite{Popovych2013}, self-organised criticality \cite{Bornholdt2003}, heterogeneous nucleation \cite{Fialkowski2023}, and others. For a more detailed overview, we refer to the recent review \cite{Berner2023} and references therein. Despite being a versatile class of models suitable for many applications, ADNs are challenging for a theoretical or numerical study due to their complexity, which is usually reflected in their high dimensionality, multistability,  or the presence of multiple time scales \cite{Kuehn2019a,Kuehn2015}. 

This paper reports on the phenomenon of recurrent adaptive chaotic clustering (RACC) in ADNs. More specifically, we describe the emergence of a robust dynamical regime in which the system exhibits ``quasi-stationary'' frequency clusters of various types. The frequency clusters exist for a relatively long time, followed by a rapid transition to another cluster or to a frequency-synchronized regime. The main difference with the previously reported recurrent switching in ADNs \cite{Thiele2023} is that in our case the switching between these quasi-stationary regimes is chaotic, and thus the length of each clustering episode cannot be predicted for a longer time. 
Looking at the observed phenomenon from the point of view of the theory of multiscale systems, the reported effect represents \textit{slow chaos}, where the chaotic dynamics is manifested on the slow timescale of the coupling weights, but the fast dynamics is regular. 
To understand this phenomenon in detail, we consider a minimal model of ADN with three oscillators and six slowly changing couplings. 

The structure of the paper is as follows. In Section \ref{sec:model} we introduce the concept of adaptive dynamical networks and describe the model under study in this paper. In Section \ref{sec:phenom} we present numerical evidence of RACC for a minimal model of three phase oscillators for specific parameter values and in Section \ref{sec:bifanalysis} we show that such behavior is robust to changes in parameters. In Section \ref{sec:lyapunov} we compute the Lyapunov exponents and show that the dynamics of the system can be chaotic and the largest Lyapunov exponent is positive for a large range of parameters. In Section \ref{sec:slowvariables} we present the new type of chaotic slow dynamics reported in this paper in a particular coupling weights subspace and demonstrate that chaotic clustering exhibits different partially synchronized states as well as complete frequency synchronization and desynchronization regimes. Section \ref{sec:concl} contains our final remarks.


\section{Adaptive dynamical network model}
\label{sec:model}

A dynamical network \cite{Strogatz_network2001,BOCCALETTI_NETWORK2006175,Mata_network2020,ZOU_NETWORK20191} is defined as a set of $N$ nodes, or vertices, and $L$ links, or edges. The topology of the connections is given by the adjacency matrix (or connectivity matrix) $A = \{a_{ij}\}$, $i,j = 1, 2, \ldots, N$, with elements equal to $1$ if the node $j$ is connected to the node $i$, and $0$ otherwise. The state of the network is given by $\vb{x} = (\vb{x}_1, \vb{x}_2, \ldots, \vb{x}_N)$, where $\vb{x}_i \in \mathbb{R}^d$ is the $d$-dimensional state variable of each individual node, and its dynamics can be written as
\begin{equation}
    \dot{\vb{x}}_i = f_i(\vb{x}_i, t) + \sum_{j=1}^Na_{ij}\Gamma_{ij}(\vb{x}_i, \vb{x}_j, t),
\end{equation}
where $f_i$ defines the local dynamics of each node, and $\Gamma_{ij}(\vb{x}_i, \vb{x}_j, t)$ is the coupling function. We can also include weights to the connections between two nodes and define a weighted dynamical network as
\begin{equation}
    \dot{\vb{x}}_i = f_i(\vb{x}_i, t) + \sum_{j=1}^N\kappa_{ij}\Gamma_{ij}(\vb{x}_i, \vb{x}_j, t),
\end{equation}
where $\kappa_{ij} \in \mathbb{R}$. By allowing these coupling weights to dynamically adapt depending on the state and the history of each node, the network structure becomes part of the temporal evolution and is not static anymore. We call such a system an adaptive dynamical network (ADN) \cite{gross2008adaptive,Maslennikov_adaptive2017,Berner2023, sawickiPerspectivesAdaptiveDynamical2023}, and define it as
\begin{align}
    \dot{\vb{x}}_i &= f_i(\vb{x}_i, t) + \sum_{j=1}^N\kappa_{ij}\Gamma_{ij}(\vb{x}_i, \vb{x}_j, t),\label{eq:networkdynamics}\\
    \dot{\kappa}_{ij} &= g_{ij}(\vb{x}_i, \vb{x}_j, t),\label{eq:adaptationrule}
\end{align}
with $g(\vb{x}_i, \vb{x}_j, t)$ being the adaptation function depending explicitly on the state of the nodes $i$ and $j$. This allows the network structure to rearrange according to the states of the nodes, which in turn are affected by this structure.

The interest in ADN of the form \eqref{eq:networkdynamics}--\eqref{eq:adaptationrule} have grown significantly over the last years \cite{Berner2023, sawickiPerspectivesAdaptiveDynamical2023}, and several forms of adaptation rules have been proposed in order to describe different dynamical systems and phenomena \cite{zhou2006dynamical, Wiedermann2015, Martens2017a, duchetMeanFieldApproximationsAdaptive2023}. One basic type of adaptation rule that has gained a lot of attention recently is the following \cite{Takaaki2009,KAS17,Berner2019,Berner2021,Thiele2023, juettner2023}
\begin{align}
    \dot{\vb{x}}_i &= f_i(\vb{x}_i) + \sum_{j = 1}^N \kappa_{ij}\Gamma_{ij}(\vb{x}_i, \vb{x}_j),\label{eq:statevariables}\\
    \dot{\kappa}_{ij} &= -\varepsilon [\kappa_{ij} + a_{ij} h_{ij}(\vb{x}_i - \vb{x}_j)],\label{eq:couplings}
\end{align}
where it is assumed that the adaptation function $h_{ij}(\vb{x}_i - \vb{x}_j)$ depends only on the difference of the corresponding state vectors, analogous to the spike-timing-dependent plasticity (STDP) \cite{Abbott2000,Caporale2008,stdp3}. The base connectivity structure is given by the matrix elements $a_{ij} \in \mathbb{R}$ and the parameter $\varepsilon > 0$ is a timescale separation parameter. 

In particular, we are interested in slow adaptation, which means $\varepsilon \ll 1$. Then, the dynamics of the nodes, Eq. \eqref{eq:statevariables}, is much faster than the dynamics of the network, and tools from the geometric singular perturbation theory can be used to study such systems \cite{Fenichel1979,Kuehn2015,krupaRelaxationOscillationCanard2001, szmolyanCanardsR32001,Wieczorek2011}.
Indeed, the slow-fast dynamics is one of the essential ingredients for the emergence of recurrent synchronization  \cite{Thiele2023} or excitability \cite{Ciszak2020a} in adaptive networks. With the slow change of the coupling weights $\kappa_{ij}$, the dynamics in the fast layer ($\mathbf{x}$) can exhibit different synchrony patterns recurrently for different weights $\kappa$.

Here, we are interested in the dynamics and synchronization of coupled nonlinear oscillator systems. Models of phase oscillators, such as the paradigmatic Kuramoto \cite{Kuramoto1984} and Kuramoto-Sakaguchi \cite{kuramotosakaguchi} models, have been of major importance in the development of the theory of synchronization, and it is widely known that we can reduce a network of coupled nonlinear oscillators to a network of phase oscillators given weak interactions \cite{reductionphase1,reductionphase2,reductionphase3}. Therefore, following Eqs. \eqref{eq:statevariables}--\eqref{eq:couplings}, we consider a network of $N$ adaptively coupled phase oscillators given by
\begin{align}
    \dot{\phi}_i &= \omega_i - \frac{1}{N}\sum_{j = 1}^N\kappa_{ij}\sin(\phi_i - \phi_j + \delta),\label{eq:model_phases}\\
    \dot{\kappa}_{ij} &= -\varepsilon\qty[\kappa_{ij} + a_{ij}\sin(\phi_i - \phi_j + \beta_{ij})],\label{eq:model_couplings}
\end{align}
where $\phi_i \in [0, 2\pi)$. The coupling and adaptation functions are homogeneously chosen, \textit{i.e.}, the same function for each pair of phase oscillators, and $\omega_i$ is the natural frequency of the $i$th oscillator. The parameter $\delta$ can be considered as a phase-lag of the interaction \cite{kuramotosakaguchi}, and $\beta_{ij}$ controls the properties of the adaptation function \cite{Berner2019,Fialkowski2023,Takaaki2009}.

In the following Section~\ref{sec:phenom}, we investigate the collective dynamics of the network \eqref{eq:model_phases}--\eqref{eq:model_couplings} with three phase oscillators, which is a minimal model exhibiting RACC. For most numerical simulations, the initial conditions are chosen randomly, with the phases, $\phi_i$, distributed uniformly in the interval $[0, 2\pi)$ and the coupling weights, $\kappa_{ij}$, in $[-1, 1]$, unless explicitly stated otherwise. All numerical integration in this paper is performed using the 4th order Runge--Kutta method implemented in Fortran.


\section{Chaotic recurrent clustering: numerical evidence}
\label{sec:phenom}

\begin{figure*}[t]
    \centering
    \includegraphics[width=0.99\linewidth]{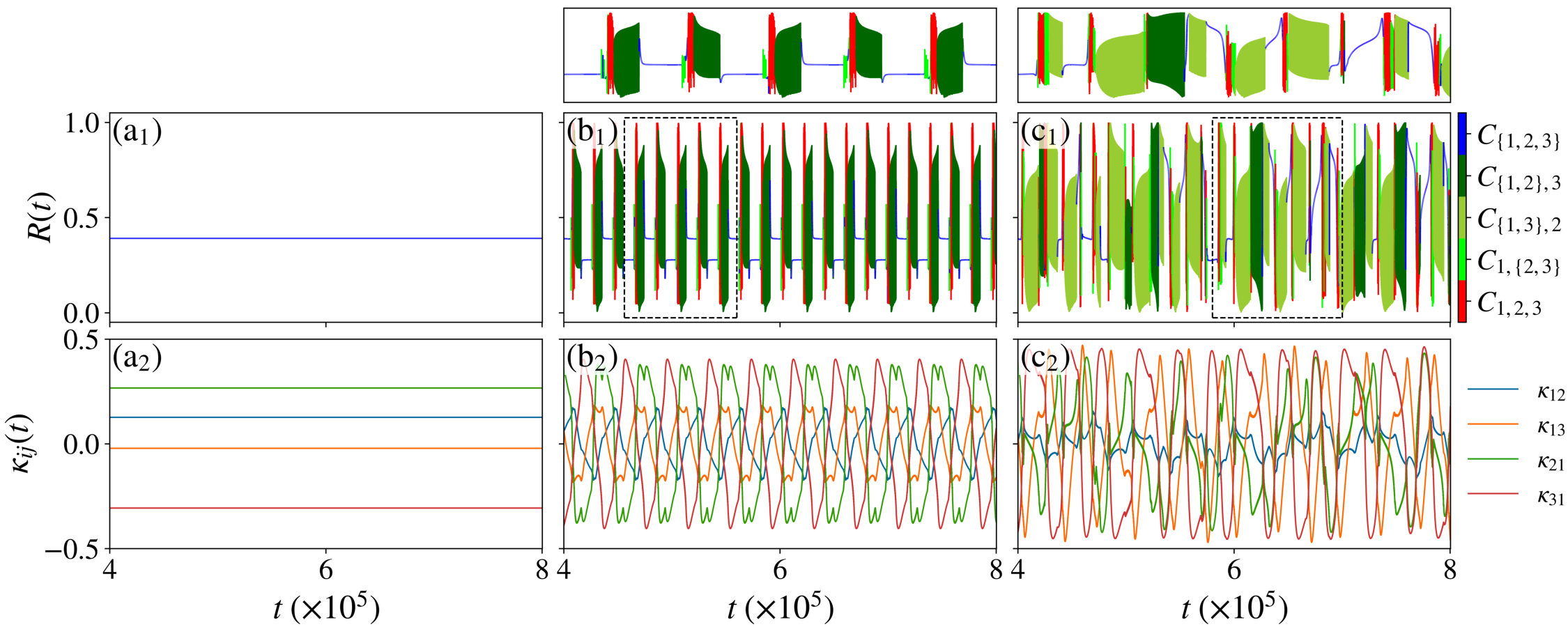}
    \caption{The order parameter, $R(t)$ [Eq. \eqref{eq:order_parameter}], and the coupling weights, $\kappa_{ij}(t) $, as a function of time for the parameters in Table \ref{tab:params} with (a) $a_{13} = 0.1$, (b) $a_{13} = 0.6$, and (c) $a_{13} = 1.1$. The top row shows magnifications in the order parameter for the corresponding regions in the dashed boxes.}
    \label{fig:examples}
\end{figure*}

To measure the collective dynamics of the network, we use an observable that provides an average of the individual nodes' dynamics, namely, the Kuramoto order parameter, $R(t)$, given by
\begin{equation}
    \label{eq:order_parameter}
    R(t) = \frac{1}{N}\abs{\sum_{j=1}^{N}e^{i\phi_j(t)}}.
\end{equation}
$R(t)$ measures the global phase synchronization of the network. If, at a certain instant of time $t$, the phases are the same, $R(t) = 1$, whereas, if the phases are spread incoherently over the interval $[0, 2\pi)$, $R(t) = 0$. Furthermore, by following the temporal evolution of $R(t)$, we can detect different synchronization states, such as frequency synchronization, which corresponds to a slowly changing $R(t)$ during some time interval, and the loss of phase relation, which corresponds to a rapidly oscillatory behavior of $R(t)$, in which we observe partial or no frequency synchronization \cite{Thiele2023}.

The phenomenon of recurrent synchronization corresponds to alternating time intervals of high activity, \textit{i.e.}, fast changes of the collective observable, and time intervals of low activity. It has been reported by Thiele \textit{et al.} \cite{Thiele2023} for two populations of Hodgkin--Huxley neurons \cite{HHneuros} with asymmetric STDP, for a reduced model of only two interacting Hodkin--Huxley neurons with asymmetric STDP, and also for two adaptively coupled phase oscillators, defined by Eqs. \eqref{eq:model_phases} and \eqref{eq:model_couplings}. Here, we study three adaptively coupled phase oscillators, and we choose the connectivity matrix elements $a_{ij}$ in such a way that there is no connection between oscillators $2$ and $3$, \textit{i.e.}, $a_{23} = a_{32} = 0$. We fix all parameters to the values shown in Table \ref{tab:params}, and change $a_{13}$, which corresponds to the amplitude of the influence of oscillator $3$ on oscillator $1$. To quantify frequency synchronization, we measure the mean phase velocity of each oscillator as
\begin{equation}
    \expval{\dot{\phi}_i} = \frac{1}{T}\int_{T_0}^{T_0 + T}\dot{\phi}_i(t)\dd{t}
\end{equation}
in windows of size $T = 300$ and say that oscillators $i$ and $j$ are frequency synchronized if $\abs{\expval{\dot{\phi}_i} - \expval{\dot{\phi}_j}} < 0.01$. If all oscillators are synchronized to each other, we have complete synchronization and one single cluster, labeled as $C_{\qty{i,j,k}}$. When only two oscillators are synchronized, we have partial synchronization and two clusters, one composed of two synchronized oscillators and one composed of a single asynchronous one. We label this state as $C_{\qty{i, j}, k}$. For the case of complete asynchrony, we have three clusters and label this state as $C_{i, j, k}$.

\begin{table}[b]
    \centering
    \caption{Parameters used in the simulations.}
    \label{tab:params}
    \begin{ruledtabular}
        \begin{tabular}{cc}
            Parameter & Value\\
            \midrule
            $\Omega_1 = \omega_1 - \omega_2$ & 0.1\\
            $\Omega_2 = \omega_1 - \omega_3$ & 0.12\\
            $\varepsilon$ & $2.0\times10^{-4}$\\
            $\delta$ & $\pi/4$\\
            $a_{12}$ & $0.375$ \\
            $a_{21}$ & $1.5$ \\
            $a_{23}$ & $0.0$ \\
            $a_{31}$ & $1.2$ \\
            $a_{32}$ & $0.0$ \\
            $\beta_{12}$ & $4\pi/3$ \\
            $\beta_{13}$ & $1 + \pi$\\
            $\beta_{21}$ & $\pi/2$ \\
            $\beta_{31}$ & $-3\pi/2 + 0.1$\\
        \end{tabular}
    \end{ruledtabular}
\end{table}


\begin{figure*}[t]
    \centering
    \includegraphics[width=0.9\linewidth]{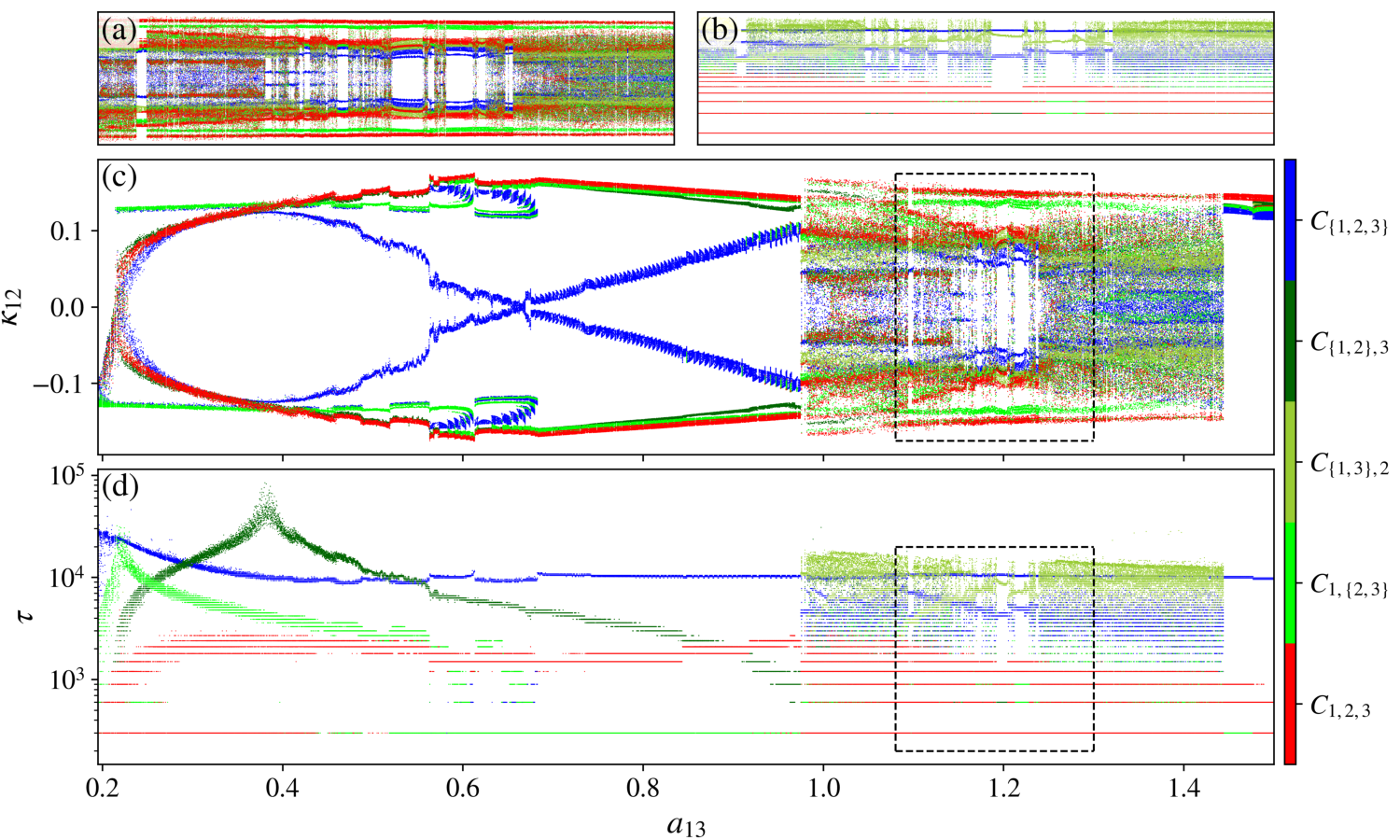}
    \caption{(a) and (c) The value of $\kappa_{12}$ when each of the clustering regimes begins, and (b) and (d) the time spent in each one of the clustering regimes as a function of $a_{13}$. Panels (a) and (b) are magnifications of the dashed black box in (c) and (d), respectively. The total integration time is $1.0\times10^6$, the transient time is $3.0\times10^5$, and the time step is $0.02$. Other parameters are in Table \ref{tab:params}.}
    \label{fig:bif_diagram}
\end{figure*}

Depending on the value of $a_{13}$, we can obtain different solutions for this model, such as the trivial fixed point solution for the coupling weights, $\kappa_{ij}$ [Fig.~\ref{fig:examples}(a)]. This corresponds to a constant value of the $R(t)$ and complete frequency synchronization, and we do not observe recurrent synchronization. As we increase the value of $a_{13}$, recurrent synchronization emerges [Fig.~\ref{fig:examples}(b)]. We clearly observe several transitions between slow changing to a fast oscillating order parameter, with slow periodic oscillations in the coupling weights during the whole depicted interval. The color scale in the first row indicates all possible synchronization states, with blue, green (all three), and red representing complete, partial, and no synchronization, respectively. In Fig.~\ref{fig:examples}(b$_1$) there is a seemingly periodic transition among these synchronization states. For larger values of $a_{13}$ the transition turns out to be rather irregular [Fig.~\ref{fig:examples}(c)], and we observe no periodicity in either the order parameter or the coupling weights. 

Interestingly, for some periods of time, oscillators 2 and 3 are synchronized even though there is no direct link between them [light green in Figs.~\ref{fig:examples}(b) and \ref{fig:examples}(c)]. Such a phenomenon is known as relay (or remote) synchronization \cite{Bergner2012,LEY18}, \textit{i.e.}, synchronization between two not directly connected oscillators in a network.

\section{Numerical bifurcation analysis of the recurrent adaptive chaotic clustering}
\label{sec:bifanalysis}

In order to study the influence of the parameters and the robustness of this newly observed phenomenon, RACC, we compute the value of one of the coupling weights, $\kappa_{12}$, when each synchronization (clustering) state begins and the time spent in each one of them as a function of the parameter $a_{13}$ (Fig.~\ref{fig:bif_diagram}). The choice of the observable $\kappa_{12}$ is not important here, only that it is one of the slow coupling variables $\kappa_{ij}$, since the fast phase variables fluctuate on a faster time-scale.  

We use random initial conditions for $a_{13} = 0$ and for each new parameter value, the initial condition corresponds to the last state of the previous value of $a_{13}$, \textit{i.e.}, we perform a brute-force numerical continuation. 
We find no recurrent clustering for values of $a_{13}$ smaller than approximately $0.2$. Beyond this value, the diagram exhibits mainly two different behaviors: ordered (periodic dynamics) and irregular (chaotic dynamics) appearance of clusters. Until $a_{13} \approx 1.0$ we observe a periodic clustering, with all cluster states present except the cluster $C_{\qty{1, 3}, 2}$ (olive green in Fig.~\ref{fig:bif_diagram}). When the influence of oscillator $3$ on oscillator $1$ increases, \textit{i.e.}, for relatively large values of $a_{13}$, we no longer observe the periodic clustering. 
Instead, the structure of the diagram is highly irregular. For some large values of $a_{13}$, there are intervals in which the diagram becomes ordered again, resembling periodic windows in classical bifurcation diagrams [Figs.~\ref{fig:bif_diagram}(a) and \ref{fig:bif_diagram}(b)]. Also, we notice the emergence of the cluster state $C_{\qty{1, 3}, 2}$. After $a_{13} \approx 1.45$ the diagram becomes regular again.

\section{Numerical Lyapunov exponents}
\label{sec:lyapunov}

The dynamics of the model can be quantified by calculating the Lyapunov exponents (LEs) \cite{Benettin1980, WOLF1985285}. In our case of only three oscillators, the dynamics of the fast system (the phases) for constant coupling weights (without adaptation) cannot be chaotic because the dynamics is effectively two-dimensional when written in terms of the phase differences $\theta_1 = \phi_1 - \phi_2$ and $\theta_2 = \phi_1 - \phi_3$. This is a result of the phase shift symmetry $\phi_i+\text{const}$ of the phase oscillator model. Also, additional zero LEs appear due to such a symmetry.

\begin{figure}[tb]
    \centering
    \includegraphics[width=0.99\linewidth]{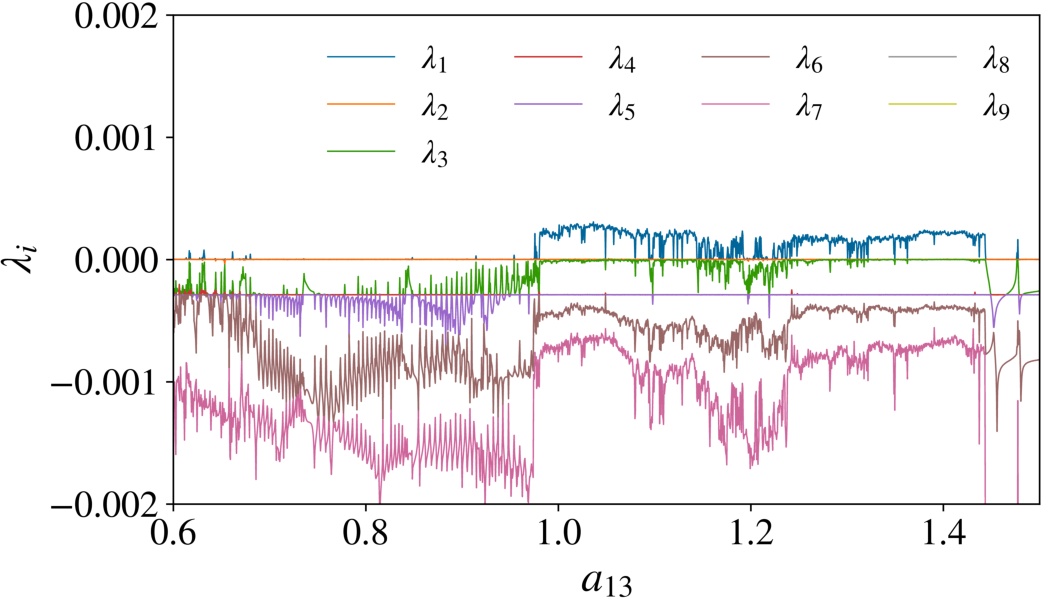}
    \caption{The Lyapunov exponents as a function of $a_{13}$. The total integration time is $1.0\times10^7$, the transient time is $3.0\times10^5$, and the time step is $0.02$. Other parameters are in Table \ref{tab:params}.}
    \label{fig:lyapunov_vs_a13}
\end{figure}

However, the whole adaptive model \eqref{eq:model_phases}--\eqref{eq:model_couplings} is high--dimensional and can exhibit chaotic solutions. In our case, we have nine equations and hence nine characteristic exponents. Our computation of the LEs follows Benettin's algorithm \cite{Benettin1980} and Wolf's code \cite{WOLF1985285}, which includes the Gram-Schimdt re-orthonormalization procedure. We compute the LEs as a function of the parameter $a_{13}$ (Fig.~\ref{fig:lyapunov_vs_a13}) using a transient time of $3.0 \times 10^5$ and total integration time of $1.0\times10^7$, with a time step of $0.02$.

\begin{figure}[tb]
    \centering
    \includegraphics[width=0.99\linewidth]{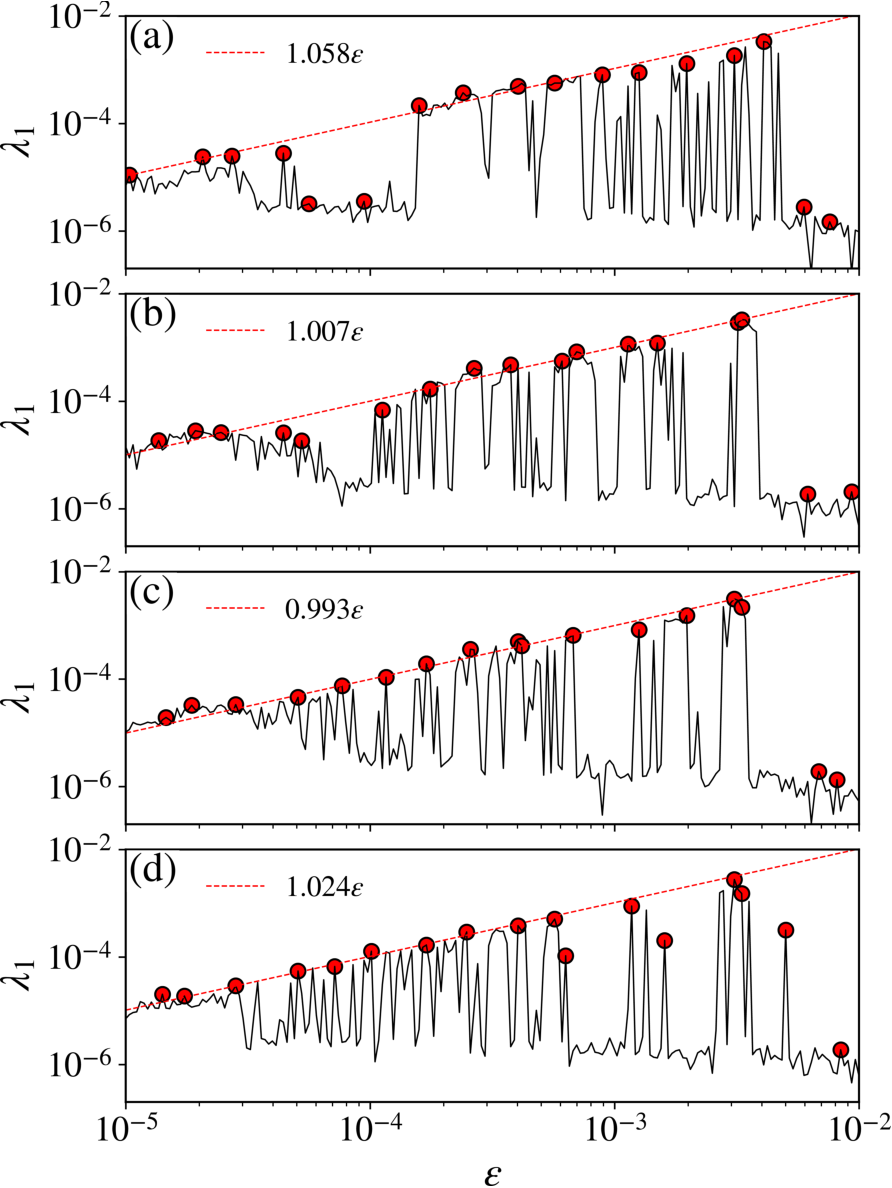}
    \caption{The largest Lyapunov exponent, $\lambda_1$, as a function of $\varepsilon$ with (a) $a_{13} = 1.0$, (b) $a_{13} = 1.1$, (c) $a_{13} = 1.2$, and (d) $a_{13} = 1.3$. The red circles were obtained by performing a max pooling in windows of 12 elements and the red dashed line represents the optimal fit based on the function $f(\varepsilon) = \mathrm{const}\times\varepsilon$ for these circles. The total integration time is $5.0\times10^6$, the transient time is $3.0\times10^5$, and the time step is $0.02$. Other parameters are in Table \ref{tab:params}.}
    \label{fig:lyapunov_vs_eps}
\end{figure}

The periodic dynamics persists for $a_{13} \lesssim 1.0$, \textit{i.e.}, there are no positive LEs, and for larger $a_{13}$ the dynamics becomes chaotic, characterized by a positive largest Lyapunov exponent, $\lambda_1 > 0$. We observe several drops of $\lambda_1$ toward zero in the ordered regions of the bifurcation diagram (Fig.~\ref{fig:bif_diagram}), indicating that these regions indeed correspond to periodic windows. Therefore, by adding a third oscillator to the network, it is possible to observe recurrent periodic synchronization as well as recurrent chaotic synchronization depending on the values of the connectivity matrix $a_{ij}$, which was not possible for the case of two oscillators \cite{Thiele2023}.

By fixing $a_{13}$ and changing $\varepsilon$ (Fig.~\ref{fig:lyapunov_vs_eps}), we find that the largest Lyapunov exponent in the chaotic regime scales as $\sim\varepsilon$, which is an indication that the chaotic dynamics occurs on the slow timescale. In the next section, we look at the dynamics of the slow variables in more detail.


\section{Dynamics of the slow variables}
\label{sec:slowvariables}

\begin{figure*}[t]
    \centering
    \includegraphics[width=0.99\linewidth]{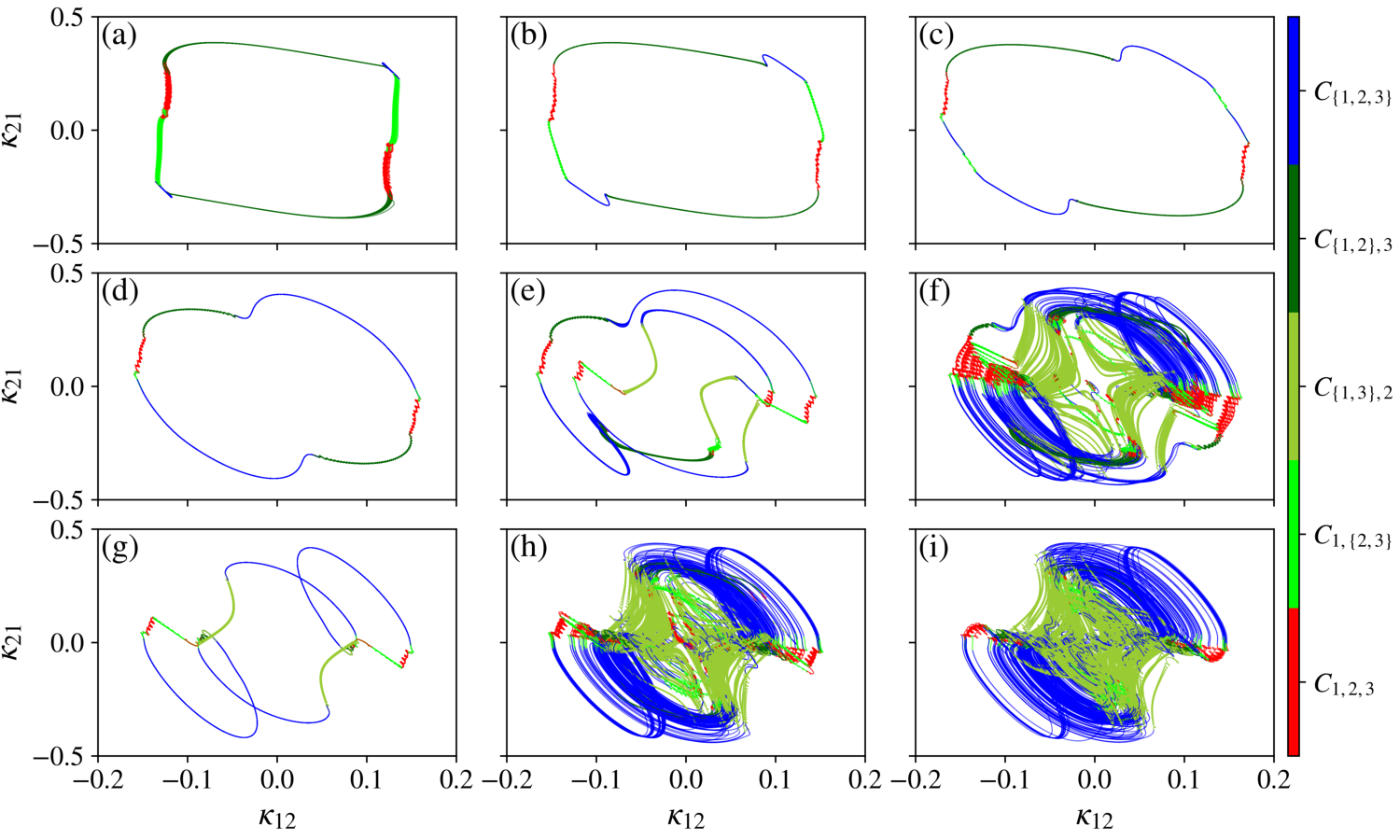}
    \caption{Projections in the $\kappa_{12}\times\kappa_{21}$ plane for (a) $a_{13} = 0.35$, (b) $a_{13} = 0.50$, (c) $a_{13} = 0.60$, (d) $a_{13} = 0.80$, (e) $a_{13} = 1.0$, (f) $a_{13} = 1.1$, (g) $a_{13} = 1.2$, (h) $a_{13} = 1.25$, and (i) $a_{13} = 1.4$. The total integration time is $5.0\times10^6$, the transient time is $3.0\times10^5$, and the time step is $0.02$. Other parameters are in Table \ref{tab:params}.}
    \label{fig:kappas_portraits}
\end{figure*}

Fig.~\ref{fig:kappas_portraits} displays the slow dynamics in the $(\kappa_{12},\kappa_{21})$ plane. When the dynamics is periodic [Figs.~\ref{fig:kappas_portraits}(a)-\ref{fig:kappas_portraits}(e), and \ref{fig:kappas_portraits}(g)], the projection into this plane shows limit cycles passing through regions of different synchronization regimes, which clarifies the behavior of the order parameter in Figs.~\ref{fig:examples}(b) and \ref{fig:examples}(c). The colors correspond to the type of clustering in Fig.~\ref{fig:examples}. In particular, blue indicates the equilibria of the fast system (slow manifolds), and the slow motions along a set of such equilibria, when all three oscillators are frequency synchronized. The different shades of green correspond to partial clusterings (see colorbar in \ref{fig:kappas_portraits}), and oscillations of the fast variables, which are averaged in the $\kappa$ dynamics. This type of dynamics, which is a more complex type of recurrent synchronization than the one reported in Ref. \cite{Thiele2023}, is shown in Figs.~\ref{fig:kappas_portraits}(f,h,i). This chaotic motion is possible because the dimensionality of the slow subsystem $\kappa_{ij}$ is 6, which gives enough dimensions for creating chaotic dynamics. In contrast, two adaptively coupled phase oscillators have only a two-dimensional slow subspace, which does not allow chaotic solutions. Similarly to the regular recurrent clustering motion, the chaotic clustering exhibits different partially synchronized states (green) as well as complete frequency synchronization (blue) and complete desynchronization (red), depending on the position in the $\kappa_{ij}$ subspace. 

\section{Conclusions}
\label{sec:concl}

In summary, by using a minimal model of three adaptively coupled phase oscillators, we have presented the phenomenon of recurrent adaptive chaotic clustering (RACC). This phenomenon is characterized by quasi-stationary frequency clusters that persist for a time interval of order $1\epsilon$, where $1/\varepsilon$ is the slow time scale of adaptation. When a quasi-stationary cluster terminates, either a different type of cluster appears, or synchrony, which is also quasi-stationary. We show that such recurrent behavior can be chaotic for a large set of parameter values. To the best of our knowledge, this phenomenon is new, and it is caused by the adaptation of the coupling weights, making it a characteristic of the class of ADNs. 

From the point of view of multiscale systems \cite{Kuehn2015,vanselowWhenVerySlow2019}, the presented phenomena provide an interesting example where the slow dynamics is chaotic while the fast (layer) dynamics is regular. We provide evidence for slow chaos by computing the maximal Lyapunov exponent and its dependence on $\varepsilon$, as well as a brute-force bifurcation diagram and a projection of the dynamics onto a plane of two slow variables. 
Any theoretical proof in this direction seems to be extremely challenging, and we would wonder if such a proof appears in the future for a possibly even simpler model of an adaptive network or some other type of slow-fast system. 

\section*{Data availability} 

The authors declare that the data supporting the findings of this study are available within the paper.
    
\section*{Declaration of competing interest}
    
The authors declare that they have no known competing financial interests or personal relationships that could have appeared to influence the work reported in this paper.

\section*{Acknowledgments}

This work was supported by the Deutsche Forschungsgemeinschaft (DFG, German Research Foundation), Project No. 411803875, and by the Brazilian agencies Coordination of Superior Level Staff Improvement (CAPES), under Grant Nos. 88887.485462/2020-00, 88881.689932/2022-01, and the São Paulo Research Foundation (FAPESP), under Grant No. 2023/08698-9.


\begin{thebibliography}{53}%
    \makeatletter
    \providecommand \@ifxundefined [1]{%
     \@ifx{#1\undefined}
    }%
    \providecommand \@ifnum [1]{%
     \ifnum #1\expandafter \@firstoftwo
     \else \expandafter \@secondoftwo
     \fi
    }%
    \providecommand \@ifx [1]{%
     \ifx #1\expandafter \@firstoftwo
     \else \expandafter \@secondoftwo
     \fi
    }%
    \providecommand \natexlab [1]{#1}%
    \providecommand \enquote  [1]{``#1''}%
    \providecommand \bibnamefont  [1]{#1}%
    \providecommand \bibfnamefont [1]{#1}%
    \providecommand \citenamefont [1]{#1}%
    \providecommand \href@noop [0]{\@secondoftwo}%
    \providecommand \href [0]{\begingroup \@sanitize@url \@href}%
    \providecommand \@href[1]{\@@startlink{#1}\@@href}%
    \providecommand \@@href[1]{\endgroup#1\@@endlink}%
    \providecommand \@sanitize@url [0]{\catcode `\\12\catcode `\$12\catcode
      `\&12\catcode `\#12\catcode `\^12\catcode `\_12\catcode `\%12\relax}%
    \providecommand \@@startlink[1]{}%
    \providecommand \@@endlink[0]{}%
    \providecommand \url  [0]{\begingroup\@sanitize@url \@url }%
    \providecommand \@url [1]{\endgroup\@href {#1}{\urlprefix }}%
    \providecommand \urlprefix  [0]{URL }%
    \providecommand \Eprint [0]{\href }%
    \providecommand \doibase [0]{http://dx.doi.org/}%
    \providecommand \selectlanguage [0]{\@gobble}%
    \providecommand \bibinfo  [0]{\@secondoftwo}%
    \providecommand \bibfield  [0]{\@secondoftwo}%
    \providecommand \translation [1]{[#1]}%
    \providecommand \BibitemOpen [0]{}%
    \providecommand \bibitemStop [0]{}%
    \providecommand \bibitemNoStop [0]{.\EOS\space}%
    \providecommand \EOS [0]{\spacefactor3000\relax}%
    \providecommand \BibitemShut  [1]{\csname bibitem#1\endcsname}%
    \let\auto@bib@innerbib\@empty
    \bibitem [{\citenamefont {Abbott}\ and\ \citenamefont
      {Nelson}(2000)}]{Abbott2000}%
      \BibitemOpen
      \bibfield  {author} {\bibinfo {author} {\bibfnamefont {L.~F.}\ \bibnamefont
      {Abbott}}\ and\ \bibinfo {author} {\bibfnamefont {S.~B.}\ \bibnamefont
      {Nelson}},\ }\bibfield  {title} {\enquote {\bibinfo {title} {Synaptic
      plasticity: {Taming} the beast},}\ }\href {\doibase 10.1038/81453} {\bibfield
       {journal} {\bibinfo  {journal} {Nature Neuroscience}\ }\textbf {\bibinfo
      {volume} {3}},\ \bibinfo {pages} {1178--1183} (\bibinfo {year}
      {2000})}\BibitemShut {NoStop}%
    \bibitem [{\citenamefont {Gerstner}\ \emph {et~al.}(2014)\citenamefont
      {Gerstner}, \citenamefont {Kistler}, \citenamefont {Naud},\ and\
      \citenamefont {Paninski}}]{Gerstner2014}%
      \BibitemOpen
      \bibfield  {author} {\bibinfo {author} {\bibfnamefont {W.}~\bibnamefont
      {Gerstner}}, \bibinfo {author} {\bibfnamefont {W.~M.}\ \bibnamefont
      {Kistler}}, \bibinfo {author} {\bibfnamefont {R.}~\bibnamefont {Naud}}, \
      and\ \bibinfo {author} {\bibfnamefont {L.}~\bibnamefont {Paninski}},\ }\href
      {\doibase 10.1017/CBO9781107447615} {\emph {\bibinfo {title} {Neuronal
      {Dynamics}}}}\ (\bibinfo  {publisher} {Cambridge University Press},\ \bibinfo
      {address} {Cambridge},\ \bibinfo {year} {2014})\BibitemShut {NoStop}%
    \bibitem [{\citenamefont {Popovych}, \citenamefont {Yanchuk},\ and\
      \citenamefont {Tass}(2013)}]{Popovych2013}%
      \BibitemOpen
      \bibfield  {author} {\bibinfo {author} {\bibfnamefont {O.}~\bibnamefont
      {Popovych}}, \bibinfo {author} {\bibfnamefont {S.}~\bibnamefont {Yanchuk}}, \
      and\ \bibinfo {author} {\bibfnamefont {P.}~\bibnamefont {Tass}},\ }\bibfield
      {title} {\enquote {\bibinfo {title} {Self-organized noise resistance of
      oscillatory neural networks with spike timing-dependent plasticity},}\ }\href
      {\doibase 10.1038/srep02926} {\bibfield  {journal} {\bibinfo  {journal}
      {Scientific Reports}\ }\textbf {\bibinfo {volume} {3}},\ \bibinfo {pages}
      {2926} (\bibinfo {year} {2013})}\BibitemShut {NoStop}%
    \bibitem [{\citenamefont {Gross}, \citenamefont {D’Lima},\ and\ \citenamefont
      {Blasius}(2006)}]{Gross2006}%
      \BibitemOpen
      \bibfield  {author} {\bibinfo {author} {\bibfnamefont {T.}~\bibnamefont
      {Gross}}, \bibinfo {author} {\bibfnamefont {C.~J.~D.}\ \bibnamefont
      {D’Lima}}, \ and\ \bibinfo {author} {\bibfnamefont {B.}~\bibnamefont
      {Blasius}},\ }\bibfield  {title} {\enquote {\bibinfo {title} {Epidemic
      {Dynamics} on an {Adaptive} {Network}},}\ }\href {\doibase
      10.1103/PhysRevLett.96.208701} {\bibfield  {journal} {\bibinfo  {journal}
      {Physical Review Letters}\ }\textbf {\bibinfo {volume} {96}},\ \bibinfo
      {pages} {208701} (\bibinfo {year} {2006})}\BibitemShut {NoStop}%
    \bibitem [{\citenamefont {Gross}\ and\ \citenamefont
      {Sayama}(2009)}]{Gross2009}%
      \BibitemOpen
      \bibfield  {author} {\bibinfo {author} {\bibfnamefont {T.}~\bibnamefont
      {Gross}}\ and\ \bibinfo {author} {\bibfnamefont {H.}~\bibnamefont {Sayama}},\
      }\enquote {\bibinfo {title} {Adaptive networks},}\ in\ \href {\doibase
      10.1007/978-3-642-01284-6_1} {\emph {\bibinfo {booktitle} {Adaptive Networks:
      Theory, Models and Applications}}},\ \bibinfo {editor} {edited by\ \bibinfo
      {editor} {\bibfnamefont {T.}~\bibnamefont {Gross}}\ and\ \bibinfo {editor}
      {\bibfnamefont {H.}~\bibnamefont {Sayama}}}\ (\bibinfo  {publisher} {Springer
      Berlin Heidelberg},\ \bibinfo {address} {Berlin, Heidelberg},\ \bibinfo
      {year} {2009})\ pp.\ \bibinfo {pages} {1--8}\BibitemShut {NoStop}%
    \bibitem [{\citenamefont {Horstmeyer}\ and\ \citenamefont
      {Kuehn}(2020)}]{Horstmeyer2020}%
      \BibitemOpen
      \bibfield  {author} {\bibinfo {author} {\bibfnamefont {L.}~\bibnamefont
      {Horstmeyer}}\ and\ \bibinfo {author} {\bibfnamefont {C.}~\bibnamefont
      {Kuehn}},\ }\bibfield  {title} {\enquote {\bibinfo {title} {Adaptive voter
      model on simplicial complexes},}\ }\href {\doibase
      10.1103/PhysRevE.101.022305} {\bibfield  {journal} {\bibinfo  {journal}
      {Physical Review E}\ }\textbf {\bibinfo {volume} {101}},\ \bibinfo {pages}
      {022305} (\bibinfo {year} {2020})}\BibitemShut {NoStop}%
    \bibitem [{\citenamefont
      {Schweitzer}(2021)}]{schweitzerSocialPercolationRevisited2021}%
      \BibitemOpen
      \bibfield  {author} {\bibinfo {author} {\bibfnamefont {F.}~\bibnamefont
      {Schweitzer}},\ }\bibfield  {title} {\enquote {\bibinfo {title} {Social
      percolation revisited: {From} 2d lattices to adaptive networks},}\ }\href
      {\doibase 10.1016/j.physa.2020.125687} {\bibfield  {journal} {\bibinfo
      {journal} {Physica A: Statistical Mechanics and its Applications}\ }\textbf
      {\bibinfo {volume} {570}},\ \bibinfo {pages} {125687} (\bibinfo {year}
      {2021})}\BibitemShut {NoStop}%
    \bibitem [{\citenamefont {Martens}\ and\ \citenamefont
      {Klemm}(2019)}]{Martens2019}%
      \BibitemOpen
      \bibfield  {author} {\bibinfo {author} {\bibfnamefont {E.~A.}\ \bibnamefont
      {Martens}}\ and\ \bibinfo {author} {\bibfnamefont {K.}~\bibnamefont
      {Klemm}},\ }\bibfield  {title} {\enquote {\bibinfo {title} {Cyclic structure
      induced by load fluctuations in adaptive transportation networks},}\ }in\
      \href {\doibase 10.1007/978-3-030-27550-1_19} {\emph {\bibinfo {booktitle}
      {Progress in Industrial Mathematics at ECMI 2018}}},\ \bibinfo {editor}
      {edited by\ \bibinfo {editor} {\bibfnamefont {I.}~\bibnamefont {Farag{\'o}}},
      \bibinfo {editor} {\bibfnamefont {F.}~\bibnamefont {Izs{\'a}k}}, \ and\
      \bibinfo {editor} {\bibfnamefont {P.~L.}\ \bibnamefont {Simon}}}\ (\bibinfo
      {publisher} {Springer International Publishing},\ \bibinfo {address} {Cham},\
      \bibinfo {year} {2019})\ pp.\ \bibinfo {pages} {147--155}\BibitemShut
      {NoStop}%
    \bibitem [{\citenamefont {Pikovsky}\ and\ \citenamefont
      {Maistrenko}(2003)}]{Pikovsky2003a}%
      \BibitemOpen
      \bibinfo {editor} {\bibfnamefont {A.}~\bibnamefont {Pikovsky}}\ and\ \bibinfo
      {editor} {\bibfnamefont {Y.}~\bibnamefont {Maistrenko}},\ eds.,\ \href@noop
      {} {\emph {\bibinfo {title} {Synchronization: {T}heory and {Application}}}},\
      Vol.\ \bibinfo {volume} {109}\ (\bibinfo  {publisher} {Kluwer Academic
      Publishers},\ \bibinfo {year} {2003})\BibitemShut {NoStop}%
    \bibitem [{\citenamefont {Pikovsky}, \citenamefont {Rosenblum},\ and\
      \citenamefont {Kurths}(2001)}]{reductionphase2}%
      \BibitemOpen
      \bibfield  {author} {\bibinfo {author} {\bibfnamefont {A.}~\bibnamefont
      {Pikovsky}}, \bibinfo {author} {\bibfnamefont {M.}~\bibnamefont {Rosenblum}},
      \ and\ \bibinfo {author} {\bibfnamefont {J.}~\bibnamefont {Kurths}},\ }\href
      {https://books.google.com.br/books?id=FuIv845q3QUC} {\emph {\bibinfo {title}
      {Synchronization: A Universal Concept in Nonlinear Sciences}}},\ Cambridge
      Nonlinear Science Series\ (\bibinfo  {publisher} {Cambridge University
      Press},\ \bibinfo {year} {2001})\BibitemShut {NoStop}%
    \bibitem [{\citenamefont {Berner}\ \emph {et~al.}(2023)\citenamefont {Berner},
      \citenamefont {Gross}, \citenamefont {Kuehn}, \citenamefont {Kurths},\ and\
      \citenamefont {Yanchuk}}]{Berner2023}%
      \BibitemOpen
      \bibfield  {author} {\bibinfo {author} {\bibfnamefont {R.}~\bibnamefont
      {Berner}}, \bibinfo {author} {\bibfnamefont {T.}~\bibnamefont {Gross}},
      \bibinfo {author} {\bibfnamefont {C.}~\bibnamefont {Kuehn}}, \bibinfo
      {author} {\bibfnamefont {J.}~\bibnamefont {Kurths}}, \ and\ \bibinfo {author}
      {\bibfnamefont {S.}~\bibnamefont {Yanchuk}},\ }\bibfield  {title} {\enquote
      {\bibinfo {title} {Adaptive dynamical networks},}\ }\href {\doibase
      https://doi.org/10.1016/j.physrep.2023.08.001} {\bibfield  {journal}
      {\bibinfo  {journal} {Physics Reports}\ }\textbf {\bibinfo {volume} {1031}},\
      \bibinfo {pages} {1--59} (\bibinfo {year} {2023})}\BibitemShut {NoStop}%
    \bibitem [{\citenamefont {Gross}\ and\ \citenamefont
      {Blasius}(2008)}]{gross2008adaptive}%
      \BibitemOpen
      \bibfield  {author} {\bibinfo {author} {\bibfnamefont {T.}~\bibnamefont
      {Gross}}\ and\ \bibinfo {author} {\bibfnamefont {B.}~\bibnamefont
      {Blasius}},\ }\bibfield  {title} {\enquote {\bibinfo {title} {Adaptive
      coevolutionary networks: a review},}\ }\href {\doibase
      10.1098/rsif.2007.1229} {\bibfield  {journal} {\bibinfo  {journal} {Journal
      of The Royal Society Interface}\ }\textbf {\bibinfo {volume} {5}},\ \bibinfo
      {pages} {259--271} (\bibinfo {year} {2008})}\BibitemShut {NoStop}%
    \bibitem [{\citenamefont {Caporale}\ and\ \citenamefont
      {Dan}(2008)}]{Caporale2008}%
      \BibitemOpen
      \bibfield  {author} {\bibinfo {author} {\bibfnamefont {N.}~\bibnamefont
      {Caporale}}\ and\ \bibinfo {author} {\bibfnamefont {Y.}~\bibnamefont {Dan}},\
      }\bibfield  {title} {\enquote {\bibinfo {title} {Spike timing-dependent
      plasticity: {A} {Hebbian} learning rule},}\ }\href {\doibase
      10.1146/annurev.neuro.31.060407.125639} {\bibfield  {journal} {\bibinfo
      {journal} {Annu. Rev. Neurosci.}\ }\textbf {\bibinfo {volume} {31}},\
      \bibinfo {pages} {25--46} (\bibinfo {year} {2008})}\BibitemShut {NoStop}%
    \bibitem [{\citenamefont {Kasatkin}\ \emph {et~al.}(2017)\citenamefont
      {Kasatkin}, \citenamefont {Yanchuk}, \citenamefont {Schöll},\ and\
      \citenamefont {Nekorkin}}]{KAS17}%
      \BibitemOpen
      \bibfield  {author} {\bibinfo {author} {\bibfnamefont {D.~V.}\ \bibnamefont
      {Kasatkin}}, \bibinfo {author} {\bibfnamefont {S.}~\bibnamefont {Yanchuk}},
      \bibinfo {author} {\bibfnamefont {E.}~\bibnamefont {Schöll}}, \ and\
      \bibinfo {author} {\bibfnamefont {V.~I.}\ \bibnamefont {Nekorkin}},\
      }\bibfield  {title} {\enquote {\bibinfo {title} {Self-organized emergence of
      multilayer structure and chimera states in dynamical networks with adaptive
      couplings},}\ }\href {\doibase 10.1103/PhysRevE.96.062211} {\bibfield
      {journal} {\bibinfo  {journal} {Physical Review E}\ }\textbf {\bibinfo
      {volume} {96}},\ \bibinfo {pages} {62211} (\bibinfo {year}
      {2017})}\BibitemShut {NoStop}%
    \bibitem [{\citenamefont {Berner}, \citenamefont {Schöll},\ and\ \citenamefont
      {Yanchuk}(2019)}]{Berner2019}%
      \BibitemOpen
      \bibfield  {author} {\bibinfo {author} {\bibfnamefont {R.}~\bibnamefont
      {Berner}}, \bibinfo {author} {\bibfnamefont {E.}~\bibnamefont {Schöll}}, \
      and\ \bibinfo {author} {\bibfnamefont {S.}~\bibnamefont {Yanchuk}},\
      }\bibfield  {title} {\enquote {\bibinfo {title} {Multiclusters in networks of
      adaptively coupled phase oscillators},}\ }\href {\doibase 10.1137/18M1210150}
      {\bibfield  {journal} {\bibinfo  {journal} {SIAM Journal on Applied Dynamical
      Systems}\ }\textbf {\bibinfo {volume} {18}},\ \bibinfo {pages} {2227--2266}
      (\bibinfo {year} {2019})}\BibitemShut {NoStop}%
    \bibitem [{\citenamefont {Feketa}, \citenamefont {Schaum},\ and\ \citenamefont
      {Meurer}(2021)}]{Feketa2021}%
      \BibitemOpen
      \bibfield  {author} {\bibinfo {author} {\bibfnamefont {P.}~\bibnamefont
      {Feketa}}, \bibinfo {author} {\bibfnamefont {A.}~\bibnamefont {Schaum}}, \
      and\ \bibinfo {author} {\bibfnamefont {T.}~\bibnamefont {Meurer}},\
      }\bibfield  {title} {\enquote {\bibinfo {title} {Synchronization and
      multicluster capabilities of oscillatory networks with adaptive coupling},}\
      }\href {\doibase 10.1109/TAC.2020.3012528} {\bibfield  {journal} {\bibinfo
      {journal} {IEEE Transactions on Automatic Control}\ }\textbf {\bibinfo
      {volume} {66}},\ \bibinfo {pages} {3084--3096} (\bibinfo {year}
      {2021})}\BibitemShut {NoStop}%
    \bibitem [{\citenamefont {Popovych}, \citenamefont {Xenakis},\ and\
      \citenamefont {Tass}(2015)}]{POP15}%
      \BibitemOpen
      \bibfield  {author} {\bibinfo {author} {\bibfnamefont {O.~V.}\ \bibnamefont
      {Popovych}}, \bibinfo {author} {\bibfnamefont {M.~N.}\ \bibnamefont
      {Xenakis}}, \ and\ \bibinfo {author} {\bibfnamefont {P.~A.}\ \bibnamefont
      {Tass}},\ }\bibfield  {title} {\enquote {\bibinfo {title} {The {Spacing}
      {Principle} for {Unlearning} {Abnormal} {Neuronal} {Synchrony}},}\ }\href
      {\doibase 10.1371/journal.pone.0117205} {\bibfield  {journal} {\bibinfo
      {journal} {PLOS ONE}\ }\textbf {\bibinfo {volume} {10}},\ \bibinfo {pages}
      {e0117205} (\bibinfo {year} {2015})}\BibitemShut {NoStop}%
    \bibitem [{\citenamefont {Aoki}(2015)}]{AOK15}%
      \BibitemOpen
      \bibfield  {author} {\bibinfo {author} {\bibfnamefont {T.}~\bibnamefont
      {Aoki}},\ }\bibfield  {title} {\enquote {\bibinfo {title} {Self-organization
      of a recurrent network under ongoing synaptic plasticity},}\ }\href {\doibase
      10.1016/j.neunet.2014.05.024} {\bibfield  {journal} {\bibinfo  {journal}
      {Neural Networks}\ }\textbf {\bibinfo {volume} {62}},\ \bibinfo {pages}
      {11--19} (\bibinfo {year} {2015})}\BibitemShut {NoStop}%
    \bibitem [{\citenamefont {Röhr}\ \emph {et~al.}(2019)\citenamefont {Röhr},
      \citenamefont {Berner}, \citenamefont {Lameu}, \citenamefont {Popovych},\
      and\ \citenamefont {Yanchuk}}]{Rohr2019}%
      \BibitemOpen
      \bibfield  {author} {\bibinfo {author} {\bibfnamefont {V.}~\bibnamefont
      {Röhr}}, \bibinfo {author} {\bibfnamefont {R.}~\bibnamefont {Berner}},
      \bibinfo {author} {\bibfnamefont {E.~L.}\ \bibnamefont {Lameu}}, \bibinfo
      {author} {\bibfnamefont {O.~V.}\ \bibnamefont {Popovych}}, \ and\ \bibinfo
      {author} {\bibfnamefont {S.}~\bibnamefont {Yanchuk}},\ }\bibfield  {title}
      {\enquote {\bibinfo {title} {Frequency cluster formation and slow
      oscillations in neural populations with plasticity},}\ }\href {\doibase
      10.1371/journal.pone.0225094} {\bibfield  {journal} {\bibinfo  {journal}
      {PLOS ONE}\ }\textbf {\bibinfo {volume} {14}},\ \bibinfo {pages} {e0225094}
      (\bibinfo {year} {2019})}\BibitemShut {NoStop}%
    \bibitem [{\citenamefont {Thiele}\ \emph {et~al.}(2023)\citenamefont {Thiele},
      \citenamefont {Berner}, \citenamefont {Tass}, \citenamefont {Schöll},\ and\
      \citenamefont {Yanchuk}}]{Thiele2023}%
      \BibitemOpen
      \bibfield  {author} {\bibinfo {author} {\bibfnamefont {M.}~\bibnamefont
      {Thiele}}, \bibinfo {author} {\bibfnamefont {R.}~\bibnamefont {Berner}},
      \bibinfo {author} {\bibfnamefont {P.~A.}\ \bibnamefont {Tass}}, \bibinfo
      {author} {\bibfnamefont {E.}~\bibnamefont {Schöll}}, \ and\ \bibinfo
      {author} {\bibfnamefont {S.}~\bibnamefont {Yanchuk}},\ }\bibfield  {title}
      {\enquote {\bibinfo {title} {Asymmetric adaptivity induces recurrent
      synchronization in complex networks},}\ }\href {\doibase 10.1063/5.0128102}
      {\bibfield  {journal} {\bibinfo  {journal} {Chaos: An Interdisciplinary
      Journal of Nonlinear Science}\ }\textbf {\bibinfo {volume} {33}},\ \bibinfo
      {pages} {023123} (\bibinfo {year} {2023})}\BibitemShut {NoStop}%
    \bibitem [{\citenamefont {Bornholdt}\ and\ \citenamefont
      {R\"ohl}(2003)}]{Bornholdt2003}%
      \BibitemOpen
      \bibfield  {author} {\bibinfo {author} {\bibfnamefont {S.}~\bibnamefont
      {Bornholdt}}\ and\ \bibinfo {author} {\bibfnamefont {T.}~\bibnamefont
      {R\"ohl}},\ }\bibfield  {title} {\enquote {\bibinfo {title} {Self-organized
      critical neural networks},}\ }\href {\doibase 10.1103/PhysRevE.67.066118}
      {\bibfield  {journal} {\bibinfo  {journal} {Phys. Rev. E}\ }\textbf {\bibinfo
      {volume} {67}},\ \bibinfo {pages} {066118} (\bibinfo {year}
      {2003})}\BibitemShut {NoStop}%
    \bibitem [{\citenamefont {Fialkowski}\ \emph {et~al.}(2023)\citenamefont
      {Fialkowski}, \citenamefont {Yanchuk}, \citenamefont {Sokolov}, \citenamefont
      {Schöll}, \citenamefont {Gottwald},\ and\ \citenamefont
      {Berner}}]{Fialkowski2023}%
      \BibitemOpen
      \bibfield  {author} {\bibinfo {author} {\bibfnamefont {J.}~\bibnamefont
      {Fialkowski}}, \bibinfo {author} {\bibfnamefont {S.}~\bibnamefont {Yanchuk}},
      \bibinfo {author} {\bibfnamefont {I.~M.}\ \bibnamefont {Sokolov}}, \bibinfo
      {author} {\bibfnamefont {E.}~\bibnamefont {Schöll}}, \bibinfo {author}
      {\bibfnamefont {G.~A.}\ \bibnamefont {Gottwald}}, \ and\ \bibinfo {author}
      {\bibfnamefont {R.}~\bibnamefont {Berner}},\ }\bibfield  {title} {\enquote
      {\bibinfo {title} {Heterogeneous {Nucleation} in {Finite}-{Size} {Adaptive}
      {Dynamical} {Networks}},}\ }\href {\doibase 10.1103/PhysRevLett.130.067402}
      {\bibfield  {journal} {\bibinfo  {journal} {Physical Review Letters}\
      }\textbf {\bibinfo {volume} {130}},\ \bibinfo {pages} {067402} (\bibinfo
      {year} {2023})}\BibitemShut {NoStop}%
    \bibitem [{\citenamefont {Kuehn}(2019)}]{Kuehn2019a}%
      \BibitemOpen
      \bibfield  {author} {\bibinfo {author} {\bibfnamefont {C.}~\bibnamefont
      {Kuehn}},\ }\bibfield  {title} {\enquote {\bibinfo {title} {Multiscale
      dynamics of an adaptive catalytic network},}\ }\href {\doibase
      10.1051/mmnp/2019015} {\bibfield  {journal} {\bibinfo  {journal}
      {Mathematical Modelling of Natural Phenomena}\ }\textbf {\bibinfo {volume}
      {14}},\ \bibinfo {pages} {402} (\bibinfo {year} {2019})}\BibitemShut
      {NoStop}%
    \bibitem [{\citenamefont {Kuehn}(2015)}]{Kuehn2015}%
      \BibitemOpen
      \bibfield  {author} {\bibinfo {author} {\bibfnamefont {C.}~\bibnamefont
      {Kuehn}},\ }\href {\doibase 10.1007/978-3-319-12316-5} {\emph {\bibinfo
      {title} {Multiple {Time} {Scale} {Dynamics}}}},\ Vol.\ \bibinfo {volume}
      {191}\ (\bibinfo  {publisher} {Springer-Verlag GmbH},\ \bibinfo {year}
      {2015})\BibitemShut {NoStop}%
    \bibitem [{\citenamefont {Strogatz}(2001)}]{Strogatz_network2001}%
      \BibitemOpen
      \bibfield  {author} {\bibinfo {author} {\bibfnamefont {S.~H.}\ \bibnamefont
      {Strogatz}},\ }\bibfield  {title} {\enquote {\bibinfo {title} {Exploring
      complex networks},}\ }\href {\doibase 10.1038/35065725} {\bibfield  {journal}
      {\bibinfo  {journal} {Nature}\ }\textbf {\bibinfo {volume} {410}},\ \bibinfo
      {pages} {268--276} (\bibinfo {year} {2001})}\BibitemShut {NoStop}%
    \bibitem [{\citenamefont {Boccaletti}\ \emph {et~al.}(2006)\citenamefont
      {Boccaletti}, \citenamefont {Latora}, \citenamefont {Moreno}, \citenamefont
      {Chavez},\ and\ \citenamefont {Hwang}}]{BOCCALETTI_NETWORK2006175}%
      \BibitemOpen
      \bibfield  {author} {\bibinfo {author} {\bibfnamefont {S.}~\bibnamefont
      {Boccaletti}}, \bibinfo {author} {\bibfnamefont {V.}~\bibnamefont {Latora}},
      \bibinfo {author} {\bibfnamefont {Y.}~\bibnamefont {Moreno}}, \bibinfo
      {author} {\bibfnamefont {M.}~\bibnamefont {Chavez}}, \ and\ \bibinfo {author}
      {\bibfnamefont {D.-U.}\ \bibnamefont {Hwang}},\ }\bibfield  {title} {\enquote
      {\bibinfo {title} {Complex networks: Structure and dynamics},}\ }\href
      {\doibase https://doi.org/10.1016/j.physrep.2005.10.009} {\bibfield
      {journal} {\bibinfo  {journal} {Physics Reports}\ }\textbf {\bibinfo {volume}
      {424}},\ \bibinfo {pages} {175--308} (\bibinfo {year} {2006})}\BibitemShut
      {NoStop}%
    \bibitem [{\citenamefont {Mata}(2020)}]{Mata_network2020}%
      \BibitemOpen
      \bibfield  {author} {\bibinfo {author} {\bibfnamefont {A.~S.~d.}\
      \bibnamefont {Mata}},\ }\bibfield  {title} {\enquote {\bibinfo {title}
      {Complex networks: a mini-review},}\ }\href {\doibase
      10.1007/s13538-020-00772-9} {\bibfield  {journal} {\bibinfo  {journal}
      {Brazilian Journal of Physics}\ }\textbf {\bibinfo {volume} {50}},\ \bibinfo
      {pages} {658--672} (\bibinfo {year} {2020})}\BibitemShut {NoStop}%
    \bibitem [{\citenamefont {Zou}\ \emph {et~al.}(2019)\citenamefont {Zou},
      \citenamefont {Donner}, \citenamefont {Marwan}, \citenamefont {Donges},\ and\
      \citenamefont {Kurths}}]{ZOU_NETWORK20191}%
      \BibitemOpen
      \bibfield  {author} {\bibinfo {author} {\bibfnamefont {Y.}~\bibnamefont
      {Zou}}, \bibinfo {author} {\bibfnamefont {R.~V.}\ \bibnamefont {Donner}},
      \bibinfo {author} {\bibfnamefont {N.}~\bibnamefont {Marwan}}, \bibinfo
      {author} {\bibfnamefont {J.~F.}\ \bibnamefont {Donges}}, \ and\ \bibinfo
      {author} {\bibfnamefont {J.}~\bibnamefont {Kurths}},\ }\bibfield  {title}
      {\enquote {\bibinfo {title} {Complex network approaches to nonlinear time
      series analysis},}\ }\href {\doibase
      https://doi.org/10.1016/j.physrep.2018.10.005} {\bibfield  {journal}
      {\bibinfo  {journal} {Physics Reports}\ }\textbf {\bibinfo {volume} {787}},\
      \bibinfo {pages} {1--97} (\bibinfo {year} {2019})}\BibitemShut {NoStop}%
    \bibitem [{\citenamefont {Maslennikov}\ and\ \citenamefont
      {Nekorkin}(2017)}]{Maslennikov_adaptive2017}%
      \BibitemOpen
      \bibfield  {author} {\bibinfo {author} {\bibfnamefont {O.~V.}\ \bibnamefont
      {Maslennikov}}\ and\ \bibinfo {author} {\bibfnamefont {V.~I.}\ \bibnamefont
      {Nekorkin}},\ }\bibfield  {title} {\enquote {\bibinfo {title} {Adaptive
      dynamical networks},}\ }\href {\doibase 10.3367/UFNe.2016.10.037902}
      {\bibfield  {journal} {\bibinfo  {journal} {Physics-Uspekhi}\ }\textbf
      {\bibinfo {volume} {60}},\ \bibinfo {pages} {694} (\bibinfo {year}
      {2017})}\BibitemShut {NoStop}%
    \bibitem [{\citenamefont {Sawicki}\ \emph {et~al.}(2023)\citenamefont
      {Sawicki}, \citenamefont {Berner}, \citenamefont {Loos}, \citenamefont
      {Anvari}, \citenamefont {Bader}, \citenamefont {Barfuss}, \citenamefont
      {Botta}, \citenamefont {Brede}, \citenamefont {Franović}, \citenamefont
      {Gauthier}, \citenamefont {Goldt}, \citenamefont {Hajizadeh}, \citenamefont
      {Hövel}, \citenamefont {Karin}, \citenamefont {Lorenz-Spreen}, \citenamefont
      {Miehl}, \citenamefont {Mölter}, \citenamefont {Olmi}, \citenamefont
      {Schöll}, \citenamefont {Seif}, \citenamefont {Tass}, \citenamefont {Volpe},
      \citenamefont {Yanchuk},\ and\ \citenamefont
      {Kurths}}]{sawickiPerspectivesAdaptiveDynamical2023}%
      \BibitemOpen
      \bibfield  {author} {\bibinfo {author} {\bibfnamefont {J.}~\bibnamefont
      {Sawicki}}, \bibinfo {author} {\bibfnamefont {R.}~\bibnamefont {Berner}},
      \bibinfo {author} {\bibfnamefont {S.~A.~M.}\ \bibnamefont {Loos}}, \bibinfo
      {author} {\bibfnamefont {M.}~\bibnamefont {Anvari}}, \bibinfo {author}
      {\bibfnamefont {R.}~\bibnamefont {Bader}}, \bibinfo {author} {\bibfnamefont
      {W.}~\bibnamefont {Barfuss}}, \bibinfo {author} {\bibfnamefont
      {N.}~\bibnamefont {Botta}}, \bibinfo {author} {\bibfnamefont
      {N.}~\bibnamefont {Brede}}, \bibinfo {author} {\bibfnamefont
      {I.}~\bibnamefont {Franović}}, \bibinfo {author} {\bibfnamefont {D.~J.}\
      \bibnamefont {Gauthier}}, \bibinfo {author} {\bibfnamefont {S.}~\bibnamefont
      {Goldt}}, \bibinfo {author} {\bibfnamefont {A.}~\bibnamefont {Hajizadeh}},
      \bibinfo {author} {\bibfnamefont {P.}~\bibnamefont {Hövel}}, \bibinfo
      {author} {\bibfnamefont {O.}~\bibnamefont {Karin}}, \bibinfo {author}
      {\bibfnamefont {P.}~\bibnamefont {Lorenz-Spreen}}, \bibinfo {author}
      {\bibfnamefont {C.}~\bibnamefont {Miehl}}, \bibinfo {author} {\bibfnamefont
      {J.}~\bibnamefont {Mölter}}, \bibinfo {author} {\bibfnamefont
      {S.}~\bibnamefont {Olmi}}, \bibinfo {author} {\bibfnamefont {E.}~\bibnamefont
      {Schöll}}, \bibinfo {author} {\bibfnamefont {A.}~\bibnamefont {Seif}},
      \bibinfo {author} {\bibfnamefont {P.~A.}\ \bibnamefont {Tass}}, \bibinfo
      {author} {\bibfnamefont {G.}~\bibnamefont {Volpe}}, \bibinfo {author}
      {\bibfnamefont {S.}~\bibnamefont {Yanchuk}}, \ and\ \bibinfo {author}
      {\bibfnamefont {J.}~\bibnamefont {Kurths}},\ }\bibfield  {title} {\enquote
      {\bibinfo {title} {Perspectives on adaptive dynamical systems},}\ }\href
      {\doibase 10.1063/5.0147231} {\bibfield  {journal} {\bibinfo  {journal}
      {Chaos: An Interdisciplinary Journal of Nonlinear Science}\ }\textbf
      {\bibinfo {volume} {33}},\ \bibinfo {pages} {071501} (\bibinfo {year}
      {2023})}\BibitemShut {NoStop}%
    \bibitem [{\citenamefont {Zhou}\ and\ \citenamefont
      {Kurths}(2006)}]{zhou2006dynamical}%
      \BibitemOpen
      \bibfield  {author} {\bibinfo {author} {\bibfnamefont {C.}~\bibnamefont
      {Zhou}}\ and\ \bibinfo {author} {\bibfnamefont {J.}~\bibnamefont {Kurths}},\
      }\bibfield  {title} {\enquote {\bibinfo {title} {Dynamical weights and
      enhanced synchronization in adaptive complex networks},}\ }\href {\doibase
      10.1103/PhysRevLett.96.164102} {\bibfield  {journal} {\bibinfo  {journal}
      {Phys. Rev. Lett.}\ }\textbf {\bibinfo {volume} {96}},\ \bibinfo {pages}
      {164102} (\bibinfo {year} {2006})}\BibitemShut {NoStop}%
    \bibitem [{\citenamefont {Wiedermann}\ \emph {et~al.}(2015)\citenamefont
      {Wiedermann}, \citenamefont {Donges}, \citenamefont {Heitzig}, \citenamefont
      {Lucht},\ and\ \citenamefont {Kurths}}]{Wiedermann2015}%
      \BibitemOpen
      \bibfield  {author} {\bibinfo {author} {\bibfnamefont {M.}~\bibnamefont
      {Wiedermann}}, \bibinfo {author} {\bibfnamefont {J.~F.}\ \bibnamefont
      {Donges}}, \bibinfo {author} {\bibfnamefont {J.}~\bibnamefont {Heitzig}},
      \bibinfo {author} {\bibfnamefont {W.}~\bibnamefont {Lucht}}, \ and\ \bibinfo
      {author} {\bibfnamefont {J.}~\bibnamefont {Kurths}},\ }\bibfield  {title}
      {\enquote {\bibinfo {title} {Macroscopic description of complex adaptive
      networks coevolving with dynamic node states},}\ }\href {\doibase
      10.1103/PHYSREVE.91.052801/FIGURES/5/MEDIUM} {\bibfield  {journal} {\bibinfo
      {journal} {Physical Review E}\ }\textbf {\bibinfo {volume} {91}},\ \bibinfo
      {pages} {052801} (\bibinfo {year} {2015})}\BibitemShut {NoStop}%
    \bibitem [{\citenamefont {Martens}\ and\ \citenamefont
      {Klemm}(2017)}]{Martens2017a}%
      \BibitemOpen
      \bibfield  {author} {\bibinfo {author} {\bibfnamefont {E.~A.}\ \bibnamefont
      {Martens}}\ and\ \bibinfo {author} {\bibfnamefont {K.}~\bibnamefont
      {Klemm}},\ }\bibfield  {title} {\enquote {\bibinfo {title} {Transitions from
      trees to cycles in adaptive flow networks},}\ }\href {\doibase
      10.3389/fphy.2017.00062} {\bibfield  {journal} {\bibinfo  {journal}
      {Frontiers in Physics}\ }\textbf {\bibinfo {volume} {5}},\ \bibinfo {pages}
      {62} (\bibinfo {year} {2017})}\BibitemShut {NoStop}%
    \bibitem [{\citenamefont {Duchet}, \citenamefont {Bick},\ and\ \citenamefont
      {Byrne}(2023)}]{duchetMeanFieldApproximationsAdaptive2023}%
      \BibitemOpen
      \bibfield  {author} {\bibinfo {author} {\bibfnamefont {B.}~\bibnamefont
      {Duchet}}, \bibinfo {author} {\bibfnamefont {C.}~\bibnamefont {Bick}}, \ and\
      \bibinfo {author} {\bibfnamefont {A.}~\bibnamefont {Byrne}},\ }\bibfield
      {title} {\enquote {\bibinfo {title} {Mean-{Field} {Approximations} {With}
      {Adaptive} {Coupling} for {Networks} {With} {Spike}-{Timing}-{Dependent}
      {Plasticity}},}\ }\href {\doibase 10.1162/neco_a_01601} {\bibfield  {journal}
      {\bibinfo  {journal} {Neural Computation}\ }\textbf {\bibinfo {volume}
      {35}},\ \bibinfo {pages} {1481--1528} (\bibinfo {year} {2023})}\BibitemShut
      {NoStop}%
    \bibitem [{\citenamefont {Aoki}\ and\ \citenamefont
      {Aoyagi}(2009)}]{Takaaki2009}%
      \BibitemOpen
      \bibfield  {author} {\bibinfo {author} {\bibfnamefont {T.}~\bibnamefont
      {Aoki}}\ and\ \bibinfo {author} {\bibfnamefont {T.}~\bibnamefont {Aoyagi}},\
      }\bibfield  {title} {\enquote {\bibinfo {title} {Co-evolution of phases and
      connection strengths in a network of phase oscillators},}\ }\href {\doibase
      10.1103/PhysRevLett.102.034101} {\bibfield  {journal} {\bibinfo  {journal}
      {Phys. Rev. Lett.}\ }\textbf {\bibinfo {volume} {102}},\ \bibinfo {pages}
      {034101} (\bibinfo {year} {2009})}\BibitemShut {NoStop}%
    \bibitem [{\citenamefont {Berner}\ \emph {et~al.}(2021)\citenamefont {Berner},
      \citenamefont {Vock}, \citenamefont {Schöll},\ and\ \citenamefont
      {Yanchuk}}]{Berner2021}%
      \BibitemOpen
      \bibfield  {author} {\bibinfo {author} {\bibfnamefont {R.}~\bibnamefont
      {Berner}}, \bibinfo {author} {\bibfnamefont {S.}~\bibnamefont {Vock}},
      \bibinfo {author} {\bibfnamefont {E.}~\bibnamefont {Schöll}}, \ and\
      \bibinfo {author} {\bibfnamefont {S.}~\bibnamefont {Yanchuk}},\ }\bibfield
      {title} {\enquote {\bibinfo {title} {Desynchronization {Transitions} in
      {Adaptive} {Networks}},}\ }\href {\doibase 10.1103/PhysRevLett.126.028301}
      {\bibfield  {journal} {\bibinfo  {journal} {Physical Review Letters}\
      }\textbf {\bibinfo {volume} {126}},\ \bibinfo {pages} {028301} (\bibinfo
      {year} {2021})}\BibitemShut {NoStop}%
    \bibitem [{\citenamefont {Jüttner}\ and\ \citenamefont
      {Martens}(2023)}]{juettner2023}%
      \BibitemOpen
      \bibfield  {author} {\bibinfo {author} {\bibfnamefont {B.}~\bibnamefont
      {Jüttner}}\ and\ \bibinfo {author} {\bibfnamefont {E.~A.}\ \bibnamefont
      {Martens}},\ }\bibfield  {title} {\enquote {\bibinfo {title} {{Complex
      dynamics in adaptive phase oscillator networks}},}\ }\href {\doibase
      10.1063/5.0133190} {\bibfield  {journal} {\bibinfo  {journal} {Chaos: An
      Interdisciplinary Journal of Nonlinear Science}\ }\textbf {\bibinfo {volume}
      {33}},\ \bibinfo {pages} {053106} (\bibinfo {year} {2023})}\BibitemShut
      {NoStop}%
    \bibitem [{\citenamefont {Markram}, \citenamefont {Gerstner},\ and\
      \citenamefont {Sjöström}(2012)}]{stdp3}%
      \BibitemOpen
      \bibfield  {author} {\bibinfo {author} {\bibfnamefont {H.}~\bibnamefont
      {Markram}}, \bibinfo {author} {\bibfnamefont {W.}~\bibnamefont {Gerstner}}, \
      and\ \bibinfo {author} {\bibfnamefont {P.~J.}\ \bibnamefont {Sjöström}},\
      }\bibfield  {title} {\enquote {\bibinfo {title} {Spike-timing-dependent
      plasticity: A comprehensive overview},}\ }\href {\doibase
      10.3389/fnsyn.2012.00002} {\bibfield  {journal} {\bibinfo  {journal}
      {Frontiers in Synaptic Neuroscience}\ }\textbf {\bibinfo {volume} {4}},\
      \bibinfo {pages} {2} (\bibinfo {year} {2012})}\BibitemShut {NoStop}%
    \bibitem [{\citenamefont {Fenichel}(1979)}]{Fenichel1979}%
      \BibitemOpen
      \bibfield  {author} {\bibinfo {author} {\bibfnamefont {N.}~\bibnamefont
      {Fenichel}},\ }\bibfield  {title} {\enquote {\bibinfo {title} {Geometric
      singular perturbation theory for ordinary differential equations},}\ }\href
      {\doibase 10.1016/0022-0396(79)90152-9} {\bibfield  {journal} {\bibinfo
      {journal} {Journal of Differential Equations}\ }\textbf {\bibinfo {volume}
      {31}},\ \bibinfo {pages} {53--98} (\bibinfo {year} {1979})}\BibitemShut
      {NoStop}%
    \bibitem [{\citenamefont {Krupa}\ and\ \citenamefont
      {Szmolyan}(2001)}]{krupaRelaxationOscillationCanard2001}%
      \BibitemOpen
      \bibfield  {author} {\bibinfo {author} {\bibfnamefont {M.}~\bibnamefont
      {Krupa}}\ and\ \bibinfo {author} {\bibfnamefont {P.}~\bibnamefont
      {Szmolyan}},\ }\bibfield  {title} {\enquote {\bibinfo {title} {Relaxation
      {Oscillation} and {Canard} {Explosion}},}\ }\href {\doibase
      10.1006/JDEQ.2000.3929} {\bibfield  {journal} {\bibinfo  {journal} {Journal
      of Differential Equations}\ }\textbf {\bibinfo {volume} {174}},\ \bibinfo
      {pages} {312--368} (\bibinfo {year} {2001})}\BibitemShut {NoStop}%
    \bibitem [{\citenamefont {Szmolyan}\ and\ \citenamefont
      {Wechselberger}(2001)}]{szmolyanCanardsR32001}%
      \BibitemOpen
      \bibfield  {author} {\bibinfo {author} {\bibfnamefont {P.}~\bibnamefont
      {Szmolyan}}\ and\ \bibinfo {author} {\bibfnamefont {M.}~\bibnamefont
      {Wechselberger}},\ }\bibfield  {title} {\enquote {\bibinfo {title} {Canards
      in {R3}},}\ }\href {\doibase 10.1006/JDEQ.2001.4001} {\bibfield  {journal}
      {\bibinfo  {journal} {Journal of Differential Equations}\ }\textbf {\bibinfo
      {volume} {177}},\ \bibinfo {pages} {419--453} (\bibinfo {year}
      {2001})}\BibitemShut {NoStop}%
    \bibitem [{\citenamefont {Wieczorek}\ \emph {et~al.}(2011)\citenamefont
      {Wieczorek}, \citenamefont {Ashwin}, \citenamefont {Luke},\ and\
      \citenamefont {Cox}}]{Wieczorek2011}%
      \BibitemOpen
      \bibfield  {author} {\bibinfo {author} {\bibfnamefont {S.}~\bibnamefont
      {Wieczorek}}, \bibinfo {author} {\bibfnamefont {P.}~\bibnamefont {Ashwin}},
      \bibinfo {author} {\bibfnamefont {C.~M.}\ \bibnamefont {Luke}}, \ and\
      \bibinfo {author} {\bibfnamefont {P.~M.}\ \bibnamefont {Cox}},\ }\bibfield
      {title} {\enquote {\bibinfo {title} {Excitability in ramped systems: {The}
      compost-bomb instability},}\ }\href {\doibase 10.1098/rspa.2010.0485}
      {\bibfield  {journal} {\bibinfo  {journal} {Proceedings of the Royal Society
      A: Mathematical, Physical and Engineering Sciences}\ }\textbf {\bibinfo
      {volume} {467}},\ \bibinfo {pages} {1243--1269} (\bibinfo {year}
      {2011})}\BibitemShut {NoStop}%
    \bibitem [{\citenamefont {Ciszak}\ \emph {et~al.}(2020)\citenamefont {Ciszak},
      \citenamefont {Marino}, \citenamefont {Torcini},\ and\ \citenamefont
      {Olmi}}]{Ciszak2020a}%
      \BibitemOpen
      \bibfield  {author} {\bibinfo {author} {\bibfnamefont {M.}~\bibnamefont
      {Ciszak}}, \bibinfo {author} {\bibfnamefont {F.}~\bibnamefont {Marino}},
      \bibinfo {author} {\bibfnamefont {A.}~\bibnamefont {Torcini}}, \ and\
      \bibinfo {author} {\bibfnamefont {S.}~\bibnamefont {Olmi}},\ }\bibfield
      {title} {\enquote {\bibinfo {title} {Emergent excitability in populations of
      nonexcitable units},}\ }\href {\doibase 10.1103/PhysRevE.102.050201}
      {\bibfield  {journal} {\bibinfo  {journal} {Phys. Rev. E}\ }\textbf {\bibinfo
      {volume} {102}},\ \bibinfo {pages} {050201} (\bibinfo {year}
      {2020})}\BibitemShut {NoStop}%
    \bibitem [{\citenamefont {Kuramoto}(1984)}]{Kuramoto1984}%
      \BibitemOpen
      \bibfield  {author} {\bibinfo {author} {\bibfnamefont {Y.}~\bibnamefont
      {Kuramoto}},\ }\href {\doibase 10.1007/978-3-642-69689-3} {\emph {\bibinfo
      {title} {{Chemical Oscillations, Waves, and Turbulence}}}},\ \bibinfo
      {series} {Springer Series in Synergetics}, Vol.~\bibinfo {volume} {19}\
      (\bibinfo  {publisher} {Springer Berlin Heidelberg},\ \bibinfo {address}
      {Berlin, Heidelberg},\ \bibinfo {year} {1984})\BibitemShut {NoStop}%
    \bibitem [{\citenamefont {Sakaguchi}\ and\ \citenamefont
      {Kuramoto}(1986)}]{kuramotosakaguchi}%
      \BibitemOpen
      \bibfield  {author} {\bibinfo {author} {\bibfnamefont {H.}~\bibnamefont
      {Sakaguchi}}\ and\ \bibinfo {author} {\bibfnamefont {Y.}~\bibnamefont
      {Kuramoto}},\ }\bibfield  {title} {\enquote {\bibinfo {title} {{A Soluble
      Active Rotater Model Showing Phase Transitions via Mutual Entertainment}},}\
      }\href {\doibase 10.1143/PTP.76.576} {\bibfield  {journal} {\bibinfo
      {journal} {Progress of Theoretical Physics}\ }\textbf {\bibinfo {volume}
      {76}},\ \bibinfo {pages} {576--581} (\bibinfo {year} {1986})}\BibitemShut
      {NoStop}%
    \bibitem [{\citenamefont {Hoppensteadt}\ and\ \citenamefont
      {Izhikevich}(1997)}]{reductionphase1}%
      \BibitemOpen
      \bibfield  {author} {\bibinfo {author} {\bibfnamefont {F.~C.}\ \bibnamefont
      {Hoppensteadt}}\ and\ \bibinfo {author} {\bibfnamefont {E.~M.}\ \bibnamefont
      {Izhikevich}},\ }\href {\doibase 10.1007/978-1-4612-1828-9} {\emph {\bibinfo
      {title} {Neuron}}},\ \bibinfo {series} {Applied Mathematical Sciences}, Vol.\
      \bibinfo {volume} {126}\ (\bibinfo  {publisher} {Springer New York},\
      \bibinfo {address} {New York, NY},\ \bibinfo {year} {1997})\ p.\ \bibinfo
      {pages} {400}\BibitemShut {NoStop}%
    \bibitem [{\citenamefont {Pietras}\ and\ \citenamefont
      {Daffertshofer}(2019)}]{reductionphase3}%
      \BibitemOpen
      \bibfield  {author} {\bibinfo {author} {\bibfnamefont {B.}~\bibnamefont
      {Pietras}}\ and\ \bibinfo {author} {\bibfnamefont {A.}~\bibnamefont
      {Daffertshofer}},\ }\bibfield  {title} {\enquote {\bibinfo {title} {Network
      dynamics of coupled oscillators and phase reduction techniques},}\ }\href
      {\doibase https://doi.org/10.1016/j.physrep.2019.06.001} {\bibfield
      {journal} {\bibinfo  {journal} {Physics Reports}\ }\textbf {\bibinfo {volume}
      {819}},\ \bibinfo {pages} {1--105} (\bibinfo {year} {2019})}\BibitemShut
      {NoStop}%
    \bibitem [{\citenamefont {Hodgkin}\ and\ \citenamefont
      {Huxley}(1952)}]{HHneuros}%
      \BibitemOpen
      \bibfield  {author} {\bibinfo {author} {\bibfnamefont {A.~L.}\ \bibnamefont
      {Hodgkin}}\ and\ \bibinfo {author} {\bibfnamefont {A.~F.}\ \bibnamefont
      {Huxley}},\ }\bibfield  {title} {\enquote {\bibinfo {title} {A quantitative
      description of membrane current and its application to conduction and
      excitation in nerve},}\ }\href {\doibase
      https://doi.org/10.1113/jphysiol.1952.sp004764} {\bibfield  {journal}
      {\bibinfo  {journal} {The Journal of Physiology}\ }\textbf {\bibinfo {volume}
      {117}},\ \bibinfo {pages} {500--544} (\bibinfo {year} {1952})}\BibitemShut
      {NoStop}%
    \bibitem [{\citenamefont {Bergner}\ \emph {et~al.}(2012)\citenamefont
      {Bergner}, \citenamefont {Frasca}, \citenamefont {Sciuto}, \citenamefont
      {Buscarino}, \citenamefont {Ngamga}, \citenamefont {Fortuna},\ and\
      \citenamefont {Kurths}}]{Bergner2012}%
      \BibitemOpen
      \bibfield  {author} {\bibinfo {author} {\bibfnamefont {A.}~\bibnamefont
      {Bergner}}, \bibinfo {author} {\bibfnamefont {M.}~\bibnamefont {Frasca}},
      \bibinfo {author} {\bibfnamefont {G.}~\bibnamefont {Sciuto}}, \bibinfo
      {author} {\bibfnamefont {A.}~\bibnamefont {Buscarino}}, \bibinfo {author}
      {\bibfnamefont {E.~J.}\ \bibnamefont {Ngamga}}, \bibinfo {author}
      {\bibfnamefont {L.}~\bibnamefont {Fortuna}}, \ and\ \bibinfo {author}
      {\bibfnamefont {J.}~\bibnamefont {Kurths}},\ }\bibfield  {title} {\enquote
      {\bibinfo {title} {Remote synchronization in star networks},}\ }\href
      {\doibase 10.1103/PhysRevE.85.026208} {\bibfield  {journal} {\bibinfo
      {journal} {Phys. Rev. E}\ }\textbf {\bibinfo {volume} {85}},\ \bibinfo
      {pages} {026208} (\bibinfo {year} {2012})}\BibitemShut {NoStop}%
    \bibitem [{\citenamefont {Leyva}\ \emph {et~al.}(2018)\citenamefont {Leyva},
      \citenamefont {Sendiña-Nadal}, \citenamefont {Sevilla-Escoboza},
      \citenamefont {Vera-Avila}, \citenamefont {Chholak},\ and\ \citenamefont
      {Boccaletti}}]{LEY18}%
      \BibitemOpen
      \bibfield  {author} {\bibinfo {author} {\bibfnamefont {I.}~\bibnamefont
      {Leyva}}, \bibinfo {author} {\bibfnamefont {I.}~\bibnamefont
      {Sendiña-Nadal}}, \bibinfo {author} {\bibfnamefont {R.}~\bibnamefont
      {Sevilla-Escoboza}}, \bibinfo {author} {\bibfnamefont {V.~P.}\ \bibnamefont
      {Vera-Avila}}, \bibinfo {author} {\bibfnamefont {P.}~\bibnamefont {Chholak}},
      \ and\ \bibinfo {author} {\bibfnamefont {S.}~\bibnamefont {Boccaletti}},\
      }\bibfield  {title} {\enquote {\bibinfo {title} {Relay synchronization in
      multiplex networks},}\ }\href {\doibase 10.1038/s41598-018-26945-w}
      {\bibfield  {journal} {\bibinfo  {journal} {Scientific Reports}\ }\textbf
      {\bibinfo {volume} {8}},\ \bibinfo {pages} {8629} (\bibinfo {year}
      {2018})}\BibitemShut {NoStop}%
    \bibitem [{\citenamefont {Benettin}\ \emph {et~al.}(1980)\citenamefont
      {Benettin}, \citenamefont {Galgani}, \citenamefont {Giorgilli},\ and\
      \citenamefont {Strelcyn}}]{Benettin1980}%
      \BibitemOpen
      \bibfield  {author} {\bibinfo {author} {\bibfnamefont {G.}~\bibnamefont
      {Benettin}}, \bibinfo {author} {\bibfnamefont {L.}~\bibnamefont {Galgani}},
      \bibinfo {author} {\bibfnamefont {A.}~\bibnamefont {Giorgilli}}, \ and\
      \bibinfo {author} {\bibfnamefont {J.-M.}\ \bibnamefont {Strelcyn}},\
      }\bibfield  {title} {\enquote {\bibinfo {title} {Lyapunov characteristic
      exponents for smooth dynamical systems and for {H}amiltonian systems; a
      method for computing all of them. part 1: Theory},}\ }\href {\doibase
      10.1007/BF02128236} {\bibfield  {journal} {\bibinfo  {journal} {Meccanica}\
      }\textbf {\bibinfo {volume} {15}},\ \bibinfo {pages} {9--20} (\bibinfo {year}
      {1980})}\BibitemShut {NoStop}%
    \bibitem [{\citenamefont {Wolf}\ \emph {et~al.}(1985)\citenamefont {Wolf},
      \citenamefont {Swift}, \citenamefont {Swinney},\ and\ \citenamefont
      {Vastano}}]{WOLF1985285}%
      \BibitemOpen
      \bibfield  {author} {\bibinfo {author} {\bibfnamefont {A.}~\bibnamefont
      {Wolf}}, \bibinfo {author} {\bibfnamefont {J.~B.}\ \bibnamefont {Swift}},
      \bibinfo {author} {\bibfnamefont {H.~L.}\ \bibnamefont {Swinney}}, \ and\
      \bibinfo {author} {\bibfnamefont {J.~A.}\ \bibnamefont {Vastano}},\
      }\bibfield  {title} {\enquote {\bibinfo {title} {Determining {L}yapunov
      exponents from a time series},}\ }\href {\doibase
      https://doi.org/10.1016/0167-2789(85)90011-9} {\bibfield  {journal} {\bibinfo
       {journal} {Physica D: Nonlinear Phenomena}\ }\textbf {\bibinfo {volume}
      {16}},\ \bibinfo {pages} {285--317} (\bibinfo {year} {1985})}\BibitemShut
      {NoStop}%
    \bibitem [{\citenamefont {Vanselow}, \citenamefont {Wieczorek},\ and\
      \citenamefont {Feudel}(2019)}]{vanselowWhenVerySlow2019}%
      \BibitemOpen
      \bibfield  {author} {\bibinfo {author} {\bibfnamefont {A.}~\bibnamefont
      {Vanselow}}, \bibinfo {author} {\bibfnamefont {S.}~\bibnamefont {Wieczorek}},
      \ and\ \bibinfo {author} {\bibfnamefont {U.}~\bibnamefont {Feudel}},\
      }\bibfield  {title} {\enquote {\bibinfo {title} {When very slow is too fast -
      collapse of a predator-prey system},}\ }\href {\doibase
      10.1016/j.jtbi.2019.07.008} {\bibfield  {journal} {\bibinfo  {journal}
      {Journal of Theoretical Biology}\ }\textbf {\bibinfo {volume} {479}},\
      \bibinfo {pages} {64--72} (\bibinfo {year} {2019})}\BibitemShut {NoStop}%
    \end{thebibliography}

%

\end{document}